\documentstyle[bal,times,psfig]{article}
\begin{document}
%

%


\ifx \htmlstyloaded\relax  \else\let\htmlstyloaded\relax\fi

%
%
\newcommand{\htmladdnormallinkfoot}[2]{#1\footnote{#2}} 

\newcommand{\htmladdnormallink}[2]{#1}

%
\newcommand{\htmladdimg}[1]{}

%
%
%
\newcommand{\externallabels}[2]{}

%
\newcommand{\externalref}[1]{}

\makeatletter
\def\makeinnocent#1{\catcode`#1=12 }
\def\csarg#1#2{\expandafter#1\csname#2\endcsname}

\def\ThrowAwayComment#1{\begingroup
    \def\CurrentComment{#1}%
    \let\do\makeinnocent \dospecials
    \makeinnocent\^^L
    \endlinechar`\^^M \catcode`\^^M=12 \xComment}
{\catcode`\^^M=12 \endlinechar=-1 %
 \gdef\xComment#1^^M{\def\test{#1}
      \csarg\ifx{PlainEnd\CurrentComment Test}\test
          \let\html@next\endgroup
      \else \csarg\ifx{LaLaEnd\CurrentComment Test}\test
            \edef\html@next{\endgroup\noexpand\end{\CurrentComment}}
      \else \let\html@next\xComment
      \fi \fi \html@next}
}
\makeatother

\def\includecomment
 #1{\expandafter\def\csname#1\endcsname{}%
    \expandafter\def\csname end#1\endcsname{}}
\def\excludecomment
 #1{\expandafter\def\csname#1\endcsname{\ThrowAwayComment{#1}}%
    {\escapechar=-1\relax
     \csarg\xdef{PlainEnd#1Test}{\string\\end#1}%
     \csarg\xdef{LaLaEnd#1Test}{\string\\end\string\{#1\string\}}%
    }}

\excludecomment{comment}

%
%
\excludecomment{rawhtml}

%
%
\excludecomment{htmlonly}
\newcommand{\html}[1]{}

%
\newenvironment{latexonly}{}{}
\newcommand{\latex}[1]{#1}

%
%
\newcommand{\hyperref}[4]{#2\ref{#4}#3}

\newcommand{\htmlref}[2]{#1}

%
%
\newcommand{\htmlimage}[1]{}

%
\newcommand{\htmladdtonavigation}[1]{}

%
%
\title{Mobile ATM Buffer Capacity Analysis\thanks{This paper is partially 
funded by ARPA contract number J-FBI-94-223.}
}
\author{\htmladdnormallink{Stephen F. Bush\thanks{The Mathematica code for 
this paper can be found on http://www.tisl.ukans.edu/\~{}sbush.}}
{http://www.tisl.ukans.edu/\~{}sbush}, Joseph B. Evans and Victor Frost
\from
Telecommunications \& Information Sciences Laboratory\\
Department of Electrical Engineering \& Computer Science\\
University of Kansas\\
Lawrence, KS 66045-2228 \\
\email{sbush@tisl.ukans.edu}
}
\markright{S. Bush, J. Evans and V. Frost
/ Mobile ATM Buffer Capacity Analysis}
\maketitle

\begin{abstract}

This paper extends a stochastic theory for buffer fill distribution
for multiple ``on'' and ``off'' sources to a mobile environment.
Queue fill distribution is described by a set of differential equations 
assuming sources alternate asynchronously between exponentially 
distributed periods in ``on'' and ``off'' states.
This paper includes the probabilities that mobile sources have links to 
a given queue. The sources represent mobile user nodes, and the queue
represents the capacity of a switch. 
This paper presents a method of analysis  which uses mobile
parameters such as speed, call rates per unit area, cell area, and
call duration and determines queue fill distribution at the ATM cell level.
The analytic results are compared with simulation results.

\end{abstract}

%
%

\newcommand{\lr}{\lambda_{R}}
\newcommand{\lrh}{\lambda_{Rh}}
\newcommand{\lrc}{\lambda_{Rc}}
\newcommand{\lrhc}{\lambda_{Rhc}}
\newcommand{\go}{\gamma_{o}}
\newcommand{\gc}{\gamma_{c}}
\newcommand{\PB}{P_{B}}
\newcommand{\Pfh}{P_{fh}}
\newcommand{\eum}{e^{-\mu_{M}t}}
\newcommand{\Fhn}{F_{Hn}}
\newcommand{\Fhh}{F_{Hh}}
\newcommand{\ftn}{f_{T_{n}}}
\newcommand{\fth}{f_{T_{h}}}
\newcommand{\Pjone}{\frac{\lr+\lrh}{\mu_{H}}P_{j-1}}
\newcommand{\Pzero}{\left[ \sum_{k=0}^{C}\frac{ \left( \lr+\lrh \right)^{k}}{k!\mu_{H}^k}\right]^{-1}}
\newcommand{\EPj}{\frac{\left( \lr+\lrh \right)^{j}}{j!\mu_{H}^{j}}P_{0}}
\newcommand{\EPO}{\sum_{k=0}^{C} \frac{\left( \lr+\lrh \right)^{k}}{k!\mu_{H}^{k}}}
\newcommand{\PiAMS}{P_{i} (t+\Delta t, x)}
\newcommand{\Pon}{\left\{ N - (i - 1) \right\} \lambda \Delta t}
\newcommand{\Poff}{ (i + 1) \Delta t}
\newcommand{\Pnc}{\left\{ 1 - \left( (N-i) \lambda + i \right) \right\} \Delta t}
\newcommand{\Pngi}[1]{P_{j \ge {#1}}}
\newcommand{\Pbd}{
\PiAMS = \Pon P_{i-1} (t, x) \nonumber \\
+ \Poff P_{i+1} (t, x) \nonumber \\
+ \Pnc P_{i} (t, x - (i-c)\Delta t) \nonumber \\
+ O(\Delta t^2)
}
\newcommand{\Pbdmobile}{
\PiAMS = \Pon P_{i-1} (t, x) \Pngi{i-1} \nonumber \\
+ \Poff P_{i+1} (t, x) \Pngi{i+1} \nonumber \\
+ \Pnc P_{i} (t, x - (i-c)\Delta t) \Pngi{i} \nonumber \\
+ O(\Delta t^2)
}
\newcommand{\Fimobile}{\sum_{j=1}^{C} P_{j} {C \choose j} \left(\frac{\lambda}{1+\lambda}\right)^j\left(\frac{1}{1+\lambda}\right)^{C-j}}
\newcommand{\eigenvect}{\phi_{i_{mobile}} \stackrel{\rm \Delta}{=} \phi_{i}P_{j=i}}
\newcommand{\genfunc}{\Phi(1) = \sum_{i=0}^{C}\phi_{i}P_{j=i}}

\section{Introduction}

The next generation of mobile systems will most likely be B-ISDN 
compatible. This paper considers buffer fill distribution in a 
mobile ATM environment. The ATM protocol standard specifies fixed 
size 53 byte cells consisting of 48 bytes of payload and a 5 byte header.
Because the ATM cell sizes are relatively small, there will
be a moderate amount of cut-through. This means that there is an increase
in performance because  ATM cells will
be available for use immediately after being de-encapsulated from the 
wireless media carrying the ATM cells. Also, mobile systems will be able 
to take advantage
of the standardized QoS parameters as well 
as having an end-to-end standards based protocol with fixed networks.
This will lead to standards based integration with fixed networks and 
integration of voice, data, and video.

M/D/1 analysis of fixed size ATM cells provides optimistic results
because M/D/1 assumes that sources are Poisson. 
A technique which does not make that assumption provides a more 
accurate analysis and is extended in this paper to a mobile 
environment.

It appears that there has been little work done concerning the effects
of mobility on ATM. This paper will attempt to build a foundation for
analyzing mobile ATM networks by extending previous work for fixed
ATM environments. This analysis
would be useful for determining the base station queue fill distribution
and probability of cell loss in a mobile environment. It would also be
useful for simplifying mobile CBR cell simulations.

\section{Mobile Systems Analysis}

\begin{figure}[htbp]
        \centerline{\psfig{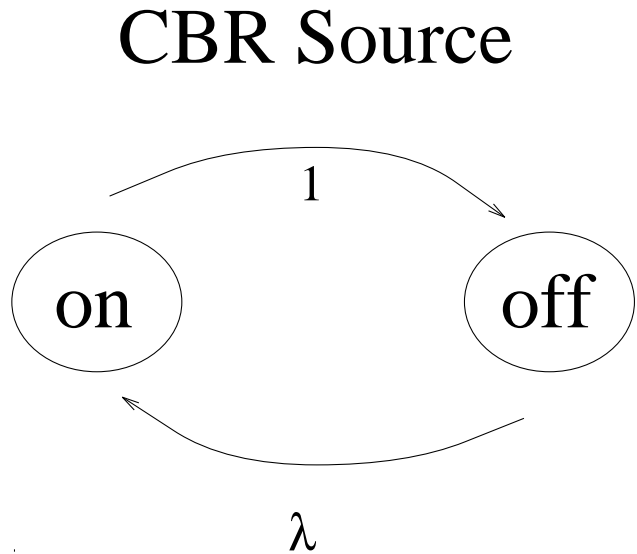}}
        \caption{Constant Bit Rate ``on''-``off'' ATM Source.}
        \label{cbr}
\end{figure}

The equilibrium buffer fill distribution can be described
by a set of differential equations assuming sources alternate
asynchronously between exponentially distributed periods in ``on'' and
``off'' states. Figure \ref{cbr} shows such a source. Note that the
``on'' to ``off'' probability is normalized to one and that the
transitions represented are intensities.

\begin{figure*}[htbp]
        \centerline{\psfig{file=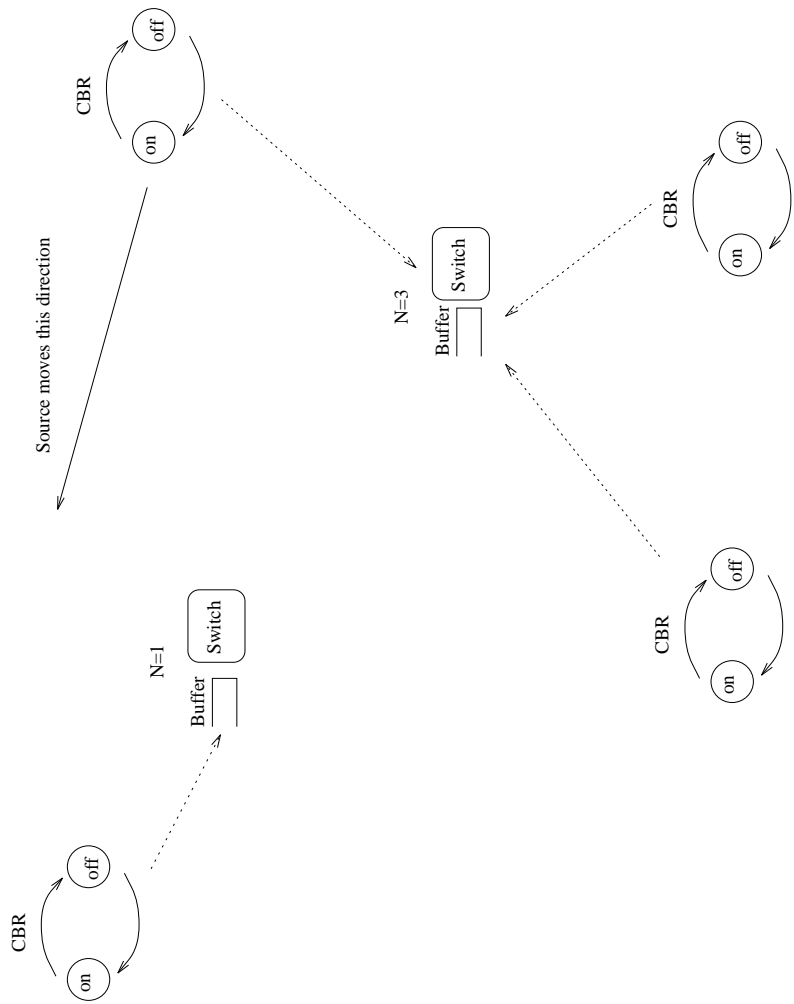,width=5.25in}}
        \caption{Mobile ATM Sources before Handoff.}
        \label{mobcbr1}
\end{figure*}

In addition, the probabilities 
that mobile sources have links to a given buffer are included. The sources 
represent 
mobile user nodes which are transmitting and receiving ATM traffic, and the 
buffer represents a switch as shown in
Figure \ref{mobcbr1}. The details of such a mobile ATM system 
implementation are described in the Rapidly Deployable Radio Networks 
(RDRN) Network Architecture \cite{BushRDRN} and \cite{BushICC96}.

\begin{figure*}[htbp]
        \centerline{\psfig{file=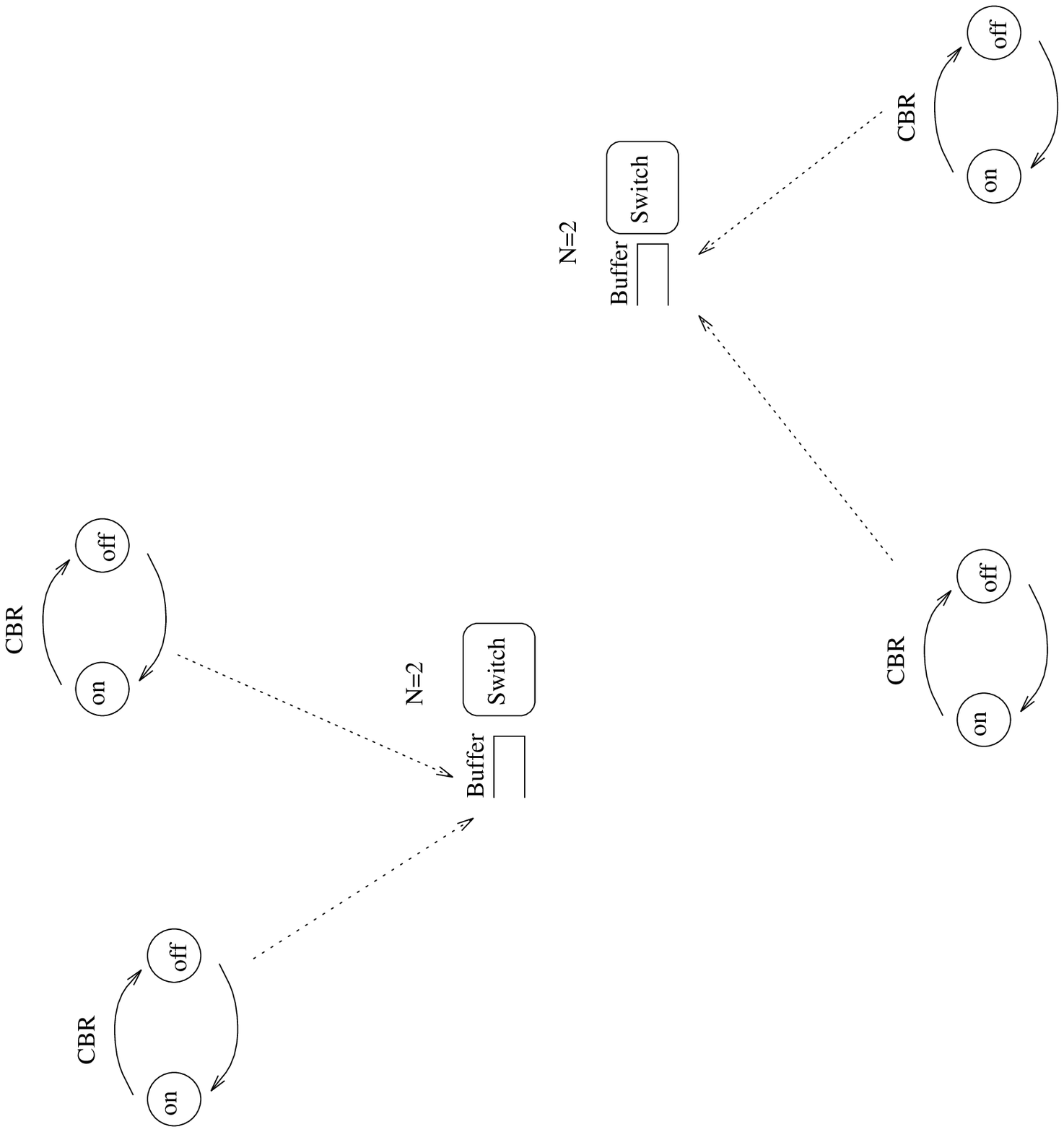,width=5.25in}}
        \caption{Mobile ATM Sources after Handoff.}
        \label{mobcbr2}
\end{figure*}

Figure \ref{mobcbr2} illustrates the handoff of a remote node from
one switch to another. Note that the ``on''-``off'' CBR Source model is
similar to the ``connected''-``disconnected'' status of the remote
nodes as they handoff from one base station to another. This observation
is used in developing the analysis for mobile ATM systems.

In a fixed network, the queue fill distribution is determined for multiple 
constant rate on-off sources. However, it assumes that the number of
sources remains constant over a sufficient period of time for the
equilibrium probabilities to be valid. There are at least two ways
of extending the analysis to a mobile cellular environment. 

Consider the ATM cell queue fill distribution at the base station.
The simplest, but least accurate extension is to determine the average 
number of channels used per cell area as $t \rightarrow \infty$. However,
there is nothing to stop mobile units from concentrating in a small
number of cell areas at some time. 
There is a hard limit on the number of sources a base station will
accept because each base station has a limited number of ports. 
Once this number is reached, further handoffs into
such a cell will cause their connections to terminate. Thus, determining
how many codes to assign to a base station is a critical design choice,
\cite{Lee}.

Note that code assignment can be dynamic, but this will not be 
considered here in order to help simplify the analysis. Also, cell areas
can be designed to overlap \cite{Lee}. Although this can increase the 
probability of interference, it has
a beneficial effect on handoff. When a mobile unit determines that a
handoff is likely to occur and the cell it will enter has no channels
available, the mobile unit can continue to use the cell within which
it currently resides, and queue the handoff to the next cell. If a channel
becomes available in the destination cell before the mobile leaves its
current cell, the handoff can take place successfully. It would be 
interesting to see the effect of queuing the handoffs. Again, in order
to keep the computation simple for this paper, this will not be 
considered.

This paper makes the simplifying assumption that each mobile unit is a 
single ATM source multiplexed at the base station. In general, each mobile
unit would be a set of sources; however, this could again be a future
extension of this paper.

\subsection{Mobile Node Analysis}

This paper makes use of the analysis and notation in \cite{Hong}.
There are two probability distributions that need to be developed:
the channel holding time and the equilibrium probability of the 
number of channels used per base switch. The channel holding time is
the probability that a particular base station's channel will be in
use at a given time, or equivalently, that a particular source still exists.
The equilibrium probability of channels in use
for a given base switch is useful in this analysis as shown later.

The first of many simplifying assumptions is that there is a known average
number of new calls per second per unit area. Let this be $\lr$
where $R$ is the radius of the particular cell area. Handoffs are attempted
at an average rate per cell, $\lrh$. The ratio of handoff attempts
to new call attempts will be $\go \stackrel{\rm \Delta}{=} \frac{\lrh}{\lr}$.

\newcommand{\Th}{T_{h}}
\newcommand{\Tm}{T_{M}}
\newcommand{\ftm}{f_{T_M}(t) = \mu_M e^{-\mu_M t}}
\newcommand{\Thn}{T_{Hn}}
\newcommand{\Thh}{T_{Hh}}
\newcommand{\Tn}{T_{n}}

Let $\PB$  be the average number of new call attempts which are blocked. Then 
new calls are accepted at an average rate of $\lrc = \lr (1-\PB)$.
Similarly, let $\Pfh$ be the average number of handoff attempts which are
blocked. Then handoff calls are accepted at a rate $\lrhc = \lrh (1-\Pfh)$.
The ratio of the average accepted handoffs to the average number of new
calls accepted is $\gc \stackrel{\rm \Delta}{=} \frac{\lrhc}{\lrc}$. The
channel holding time, $\Th$ , is a random variable defined as the time
beginning when a channel is accessed, either via a new call or handoff,
until the channel is released, via handoff or call completion. In order
to define this, another random variable, $\Tm $ is defined. $\Tm$ is
the time duration of a call, regardless of handoff or blocking.
It is simplified as an exponential with average value, $\frac{1}{\mu_M}$.
Thus the pdf is
\begin{equation}
\label{ftm}
\ftm
\end{equation}

The strategy for determining the channel holding
time distribution is to consider the time remaining for a call which
has not been handed off yet, $\Thn $, and the time remaining after
a handoff, $\Thh$. Since the call duration, $\Tm$ is memoryless,
the time remaining for a call after handoff has the same distribution
as the original call duration. Let $\Tn$ be the time the mobile
unit remains in the original cell area, and $\Th$ be the time the
mobile resides in the cell area after handoff. $\Thn$ is the minimum
of the call duration, $\Tm$, or the dwell time in the originating
cell area, $\Tn$. A similar reasoning applies to the cell area into
which a mobile unit has moved after a handoff; $\Thh$ is the minimum
of the call duration, $\Tm$, or the dwell time in the cell area after
handoff, $\Th$.

\newcommand{\Fthn}{F_{T_{Hn}}}
\newcommand{\Ftm}{F_{T_{M}}}
\newcommand{\Ftn}{F_{T_{n}}}
\newcommand{\Fth}{F_{T_{h}}}
\newcommand{\Fthh}{F_{T_{Hh}}}

Thus, 
\begin{eqnarray}
\label{FTH}
\Fthn(t) & = & \Ftm(t) + \Ftn(t)(1 - \Ftm(t)) \nonumber \\
\Fthh(t) & = & \Ftm(t) + \Fth(t)(1 - \Ftm(t)) 
\end{eqnarray}
where $(1-\Ftm(t))$ is the probability that a call does not complete
within the current cell area.
\newcommand{\FtHtwo}{\frac{\lrc}{\lrc+\lrhc}\Fthn (t)+\frac{\lrhc}{\lrc + \lrhc}\Fthh (t)}

The distribution of channel holding time in a particular cell area is 
a weighted function of the equations shown in \ref{FTH} above,
\begin{equation}
F_{T_H}(t) = \FtHtwo
\end{equation}

\newcommand{\FtHone}{1-\eum+\frac{\eum}{1+\gc}(\Ftn(t)+\gc\Fth(t))}

Substituting the values from Equation \ref{ftm},
\begin{equation}
\label{FTHone}
F_{T_H}(t) = \FtHone
\end{equation}
and differentiating to get the pdf,
\begin{eqnarray}
\label{fTH}
f_{T_H}(t) & = & \mu_{M} \eum \nonumber \\
	   &   & + \frac{\eum}{1 + \gc} \left[ \ftn (t)+ \gc \fth (t) \right]
\nonumber \\
           &   & - \frac{\eum }{1 + \gc} \left[ \Ftn (t) + \gc \Fth (t) \right]
\end{eqnarray}

To determine the equilibrium probability of the number of mobile
hosts using a base station, approximate the channel holding time as
simply an exponential distribution. The birth-death
Markov chain can be used to find the equilibrium probability of 
the number of sources in each cell area. The {\em up rates} are
$\lambda_{R} + \lambda_{Rh}$ and the {\em down rates} are multiples
of the mean channel holding time.

Putting the Markov chain in closed form,
\begin{equation}
\label{EPj}
P_j = \EPj
\end{equation}
where,
\begin{equation}
P_0 = \frac{1}{\EPO}
\end{equation}
Note that $C$ is the total number of channels for a base station and
handoffs will fail with probability $P_C$, i.e. all channels
in that cell area are currently in use.

\subsection{Mobile CBR Source Analysis}

Assume that the number of mobile hosts in a cell area is independent of
whether its CBR source is on or off. We can now modify the probability
that $i$ sources are on and the queue fill is less than $x$ by incorporating
the probability that there are at least $i$ sources, 
shown in Equation \ref{mobmod}.
\begin{equation}
\label{mobmod}
P_{i_{mobile}}(t,x) \stackrel{\rm \Delta}{=} P_{j \geq i}\ and\ P_i(t,x)
\end{equation}
$P_i(t,x)$ is the probability that at time $t$, $i$ sources are on, and
the number of items in the buffer does not exceed $x$. $P_{j \geq i}$ is the
probability that there are at least $i$ sources sending data to the buffer.
$P_{j \geq i}$ can be found from Equation \ref{EPj} as follows,
\begin{equation}
\Pngi{i} = \sum_{j=i}^C P_j
\end{equation}

The buffer fill distribution as defined in \cite{Anick1982} is
\begin{eqnarray}
\label{Pbd}
\Pbd
\end{eqnarray}

Now that the channel equilibrium probabilities have been determined,
we can account for the fact that the sources
are mobile. Since the channel equilibrium probabilities have no
dependence on time, the method of solution in \cite{Anick1982} can
be used with minor modifications,
\begin{eqnarray}
\label{Pbdmobile}
\Pbdmobile
\end{eqnarray}

From \cite{Anick1982}, $F_i(x)$ is the equilibrium probability that $i$
sources are on, and the buffer content does not exceed $x$. Thus
$F_i(\infty)$ is the probability that $i$ out of $N$ sources are
simultaneously on. In the mobile environment, this is now, 
\begin{equation}
F_i(\infty) = \Fimobile
\end{equation}

The mobile extension from Equation \ref{mobmod} carries through \cite{Anick1982}
for example, equation (13) in \cite{Anick1982} is now,
\begin{equation}
\eigenvect
\end{equation}
and
\begin{equation}
\genfunc
\end{equation}

$\phi_{i}$ is the right eigenvector of
\begin{equation}
zD\phi = M\phi
\end{equation}
where D and M are matrices used to represent the differential equation
in Equation \ref{Pbd}.

$\Phi(x)$ is the generating function of $\phi$. These values are useful
in \cite{Anick1982} for solving the equilibrium buffer fill differential
equation. The remainder of the solution is straight forward from 
\cite{Anick1982}. Thus it has been shown how an analysis of constant bit
rate on-off sources which model fixed length ATM packet sources, is 
extended to a mobile environment.

Note that the analysis uses a technique which is more accurate than
$M/D/1$ for the fixed size ATM cells, yet uses a memoryless analysis 
for the channel holding time distribution. This is a reasonable approach 
since the variable length channel hold times can be accurately modeled
by a memoryless analysis, while the $M/D/1$ analysis yields optimistic
results which can be replaced by the more accurate method in \cite{Anick1982} 
as this section has described.

\label{extension}

\subsection{Example}

The following is a simple example of the analysis using the same
parameters as the simulation in the next section. The parameters
required are: 
\begin{itemize}
\item $\lr = 0.06$ calls/sec/square mile
\item $T_{M} = 40$ secs
\item $V_{max} = 0.03$ miles/sec
\item $C = 3$ channels per base station
\end{itemize}

From the equations in the previous section, we can develop an analytical
solution for the remaining parameters. Using basic probability the integral 
of Equation \ref{fTH} should be one. Also, $\mu_{H}$ is an exponential
approximation of Equation \ref{FTHone}, which can be found by taking the 
integral of the difference of $F_{T_H}(t)$ from Equation \ref{FTHone} and 
$\mu_{H}$ and setting the result to zero,

\begin{equation}
\int_0^\infty f_{TH}(t) dt = 1
\end{equation}

\begin{equation}
\int_0^\infty \left[ F_{TH}(t) - e^{\mu_{H} t} \right] dt = 0.
\end{equation}

\begin{figure}[htbp]
        \centerline{\psfig{file=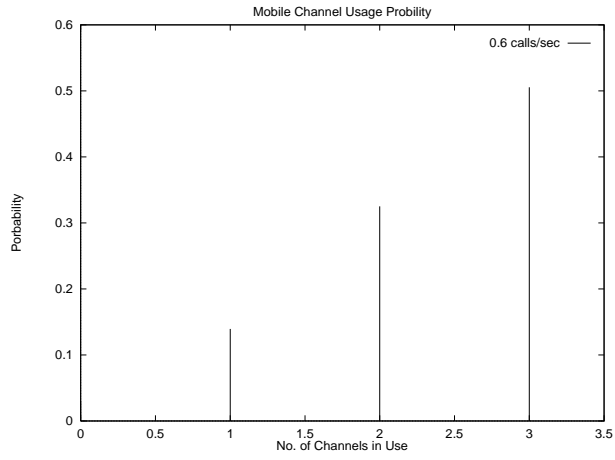,width=3.25in}}
        \caption{Channel Usage Prob. Density Function for 0.6 Call/Sec.}
        \label{Pj6_graph}
\end{figure}

\begin{figure}[htbp]
        \centerline{\psfig{file=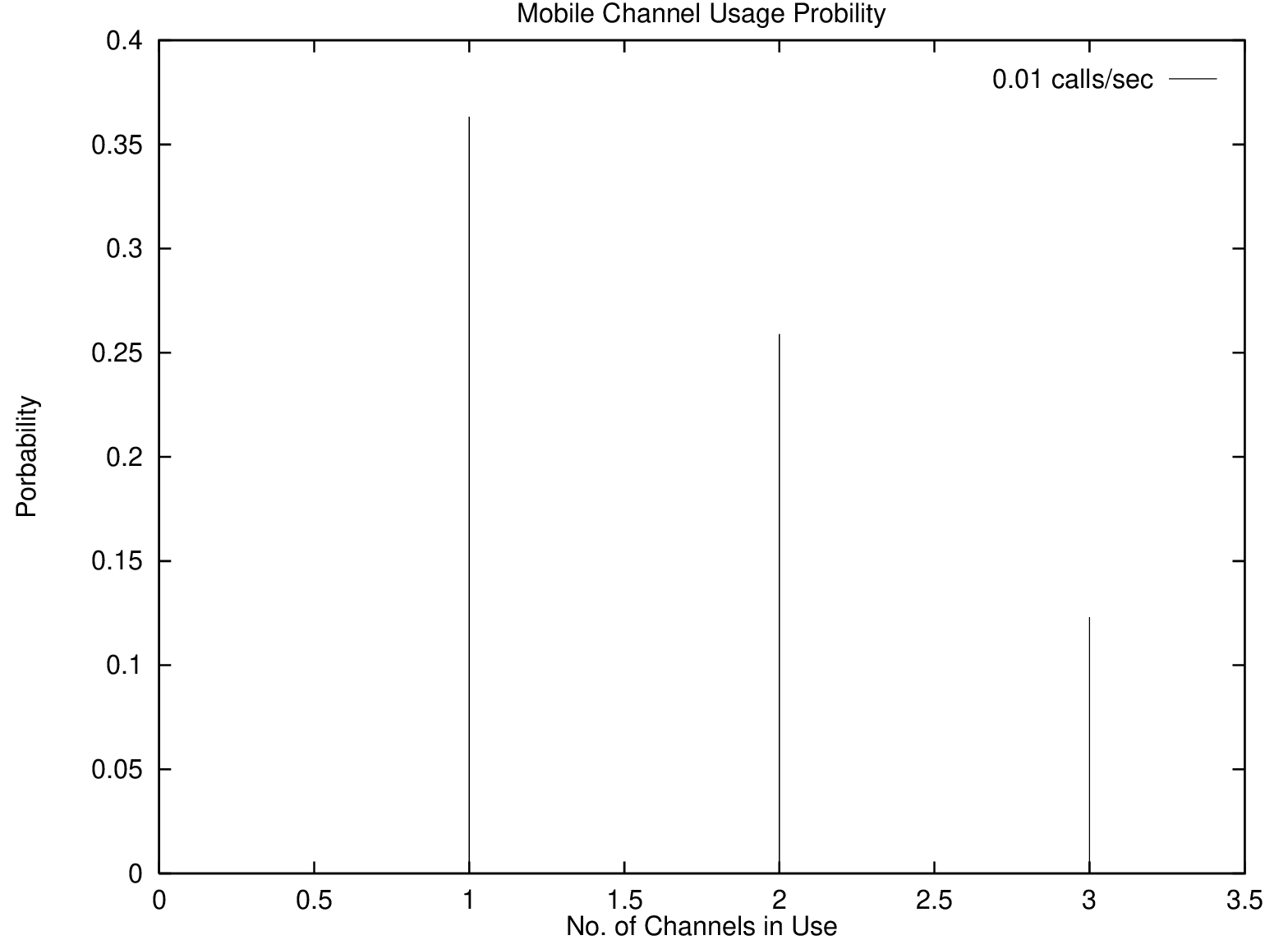,width=3.25in}}
        \caption{Channel Usage Prob. Density Function for 0.01 Call/Sec.}
        \label{Pj01_graph}
\end{figure}

This provides two equations and two unknowns which provide the solution
for $\lambda_{Rh}=2.16$ and $\mu_{H}=9.48$. These values can be
used to determine the $P_{j}$ which can then be used in our
extension of \cite{Anick1982} as discussed previously. In the graph of $P_{j}$
shown in Figures \ref{Pj6_graph}, it appears that there 
will be a high probability of blocking, since $P_{B} = P_{C} = P_{3}$.
This is compared with an arrival rate of 0.01 calls/sec/unit 
area in Figure \ref{Pj01_graph}, which has a maximum at one 
channel per station, and a lower blocking probability.

Figures \ref{Pj6_graph} and \ref{Pj01_graph} are in agreement with the 
simulation results in Figures \ref{changraph6} and \ref{changraph01}
as additional verification. 

\subsection{Simulation and Results}

\begin{figure*}[htbp]
        \centerline{\psfig{file=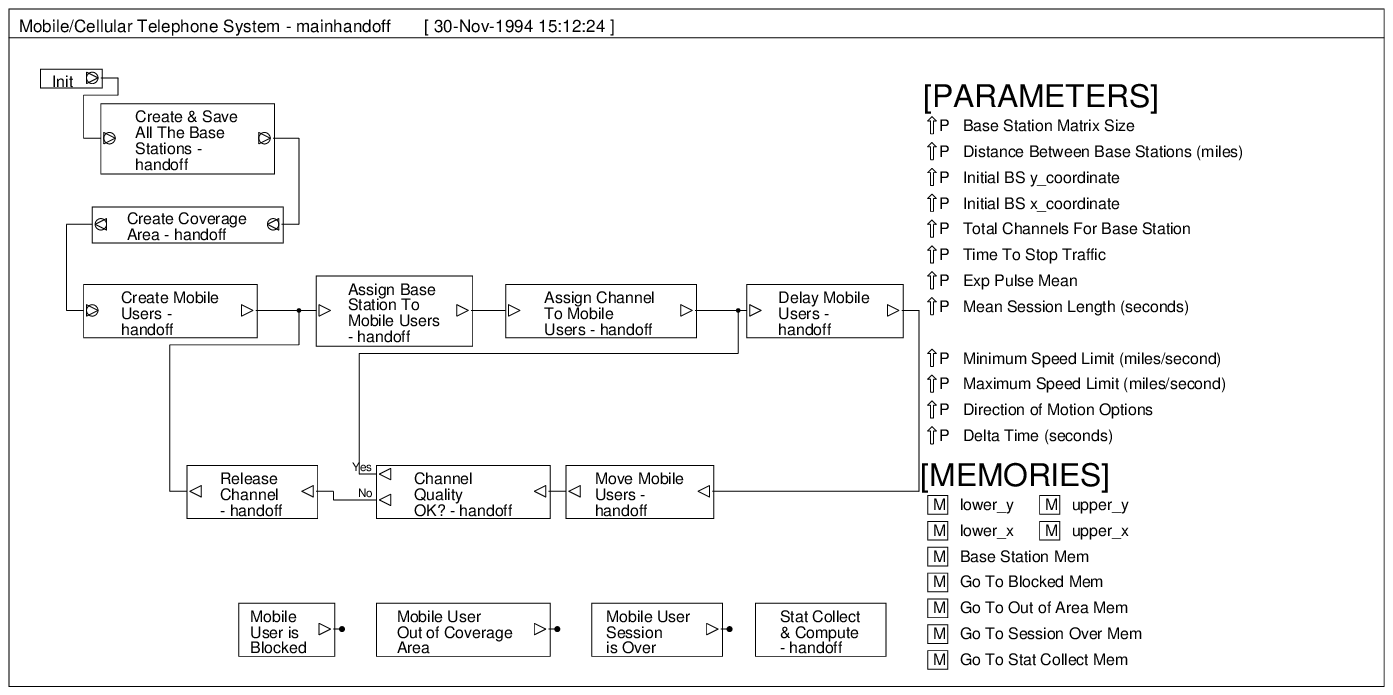,width=5.25 in}}
        \caption{Top Level Mobile System.}
        \label{mobile-cellular-telephone-system}
\end{figure*}

The mobile communications system model\footnote{A mobile cellular telephone 
system library comes with the BONeS software. As much as possible of that 
library is used as a basis 
for this simulation.} is shown in figure \ref{mobile-cellular-telephone-system}.
It is an open system; mobile hosts are generated at rate with inter-arrival
time specified by {\bf Exp Pulse Mean}, initiate a call for an average 
exponential duration specified by {\bf Mean Session Length}, and exit the 
system when either the call is complete or the mobile moves out of all cell 
areas.

The first step is to create the base stations and their cell areas.
The total number of base stations is:
\begin{equation}
{\bf Base\ Station\ Matrix\ Size}^2
\end{equation}
and they are located in a square array. They all have the same number
of channels, {\bf Total Channels for Base Station}. Each cell area can
be approximated as a circle with radius: 
\begin{equation}
\frac{{\bf Distance\ Between\ Base\ Stations}}{2}
\end{equation}

Mobile hosts are created in {\bf Create Mobile Users}. All the
mobile parameters are uniformly distributed, except the session duration
which is exponential and agrees with equation \ref{ftm} of our analysis.
New mobiles enter the system with an interarrival time of 
{\bf Exp Pulse Mean}. The mobile host will arrive at a location
uniformly distributed anywhere in the area covered by all cells. 
Since a mobile makes one call in its lifetime, a mobile host represents
a single connection. Thus, 
\begin{equation}
\lr = \frac{1}{({\bf Exp\ Pulse\ Mean})(Total\ Cell\ Area)(B)}
\end{equation}
where $B$ is the number of base stations.

The following modules act upon the mobile hosts throughout their lifetime.
The mobile hosts are assigned the nearest base station 
{\bf Assign Base Station to Mobiles Users}, and an available channel from that 
base station {\bf Assign Channel to Mobile Users}. Then all
mobile hosts dwell in their cell areas for time {\bf Delta Time} 
{\bf Delay Mobile Users}. The mobile host then moves to its next location
which may be uniformly chosen from {\bf Direction of Motion Options} and
may lead to a new cell or completely outside the cellular system
{\bf Move Mobile Users}.

After moving, the quality of signal is checked 
and if below a given criteria\footnote{In this
case, if the distance between a mobile host and its currently assigned
base station is greater than $\frac{2}{3} \sqrt{D^2-\left({\frac{D}{2}}
\right)^2}$ 
where $D$ is the distance between base stations, then
the channel quality is considered unacceptable.}
the channel is released ({\bf Release Channel}) and
the mobile host is reassigned to a new base station
({\bf Assign Base Station to Mobile Users}).

Note that the
direction of travel by a mobile user is uniformly chosen from North, 
South, East, or West. A more
sophisticated analysis of speed and direction is contained in \cite{Leung}.

\begin{figure}[htbp]
        \centerline{\psfig{file=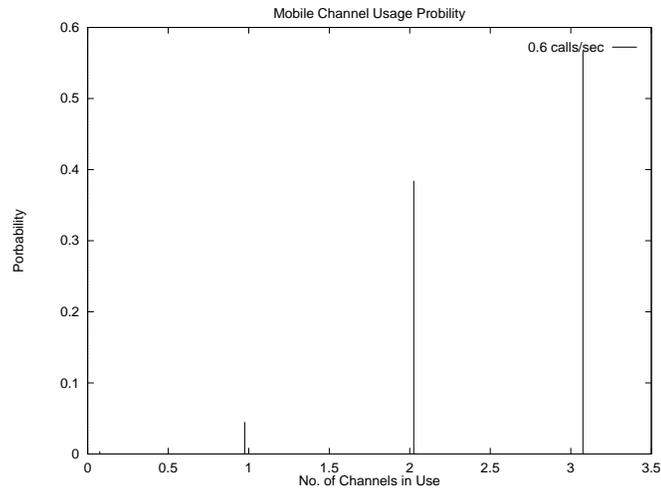,width=3.5in}}
        \caption{Channel Usage Probability for 0.6 Call/Sec.}
        \label{changraph6}
\end{figure}

\begin{figure}[htbp]
        \centerline{\psfig{file=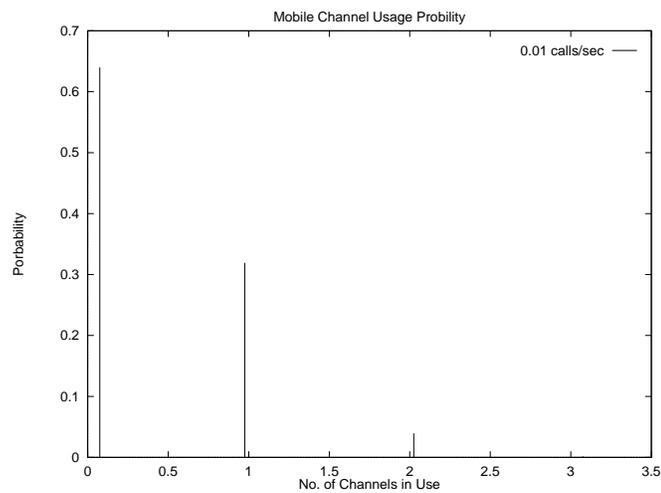,width=3.5in}}
        \caption{Channel Usage Probability for 0.01 Call/Sec.}
        \label{changraph01}
\end{figure}

The simulation results are shown in Figures \ref{changraph6} and 
\ref{changraph01}. These results agree with the results of the analysis
shown in Figures \ref{Pj6_graph} and \ref{Pj01_graph}. 

\section{Summary}

This paper attempted to extend \cite{Anick1982} to a mobile 
environment. It also presented the results of a simulation
of a mobile environment in preparation for simulating the
extension. The mobile environment adds several new dimensions
to fixed communications analysis, such as cell area, speed, 
direction, channel holding time, channels used at a base station, 
frequency of handoffs.
This simulation concentrated on finding channel hold time, $T_H$,
and the average number of channels used, $P_j$, which are required
for the extension of \cite{Anick1982}.

This paper also suggested areas for further research such as 
extending \cite{Anick1982} to mobiles which are treated as
multiple CBR sources, analyzing queued handoffs, and enhancing
the channel hold time and number of channels used by using a 
PDE for the speed and direction analysis as in \cite{Leung}.

%
%
\medskip
\bibliography{/home/bushsf/ref/mob,/home/bushsf/ref/standards}
\bibliographystyle{plain} 
\end{document}